# Multiscale modeling of thermal properties in Polyurethane incorporated with phase change materials composites: A case study

*Bokai Liu*[a], *Weizhuo Lu*[a], *Xiaoyue Hu*[b], *Chao Zhang*[c,d,e,f], *Cuixia Wang*[c,d,e,f], *Yilin Qu*[g], *Thomas Olofsson*[a]

[a] Intelligent Human-Buildings Interaction Lab (IHBI), Department of Applied Physics and Electronics, Umeå University, Håken Gullessons väg 20, 90187 Umeå, Sweden

[b] Faculty of Architecture and Urbanism, Bauhaus-Universität Weimar, 99423 Weimar, Germany

[c] Yellow River Laboratory, Zhengzhou University, Zhengzhou 450001, China

[d] Institute of Underground Engineering, Zhengzhou University, Zhengzhou 450001, China

[e] National Local Joint Engineering Laboratory of Major Infrastructure Testing and Rehabilitation Technology, Zhengzhou 450001, China

[f] Collaborative Innovation Center for disaster prevention and control of Underground Engineering jointly built by provinces and ministries, Zhengzhou, 450001, China

[g] State Key Laboratory for Strength and Vibration of Mechanical Structures, Xi'an Jiaotong University, Xi'an 710049, Shaanxi, China

**Abstract.** Polyurethane (PU) is an ideal thermal insulation material due to its excellent thermal properties. The incorporation of Phase Change Materials (PCMs) capsules into Polyurethane (PU) has been shown to be effective in building envelopes. This design can significantly increase the stability of the indoor thermal environment and reduce the fluctuation of indoor air temperature. We develop a multiscale model of a PU-PCM foam composite and study the thermal conductivity of this material. Later, the design of materials can be optimized by obtaining thermal conductivity. We conduct a case study based on the performance of this optimized material to fully consider the thermal comfort of the occupants of a building envelope with the application of PU-PCMs composites in a single room. At the same time, we also predict the energy consumption of this case. All the outcomes show that this design is promising, enabling the passive design of building energy and significantly improving occupants' comfort.

**Keywords.** Polyurethane (PU), Phase Change Materials (PCMs), Thermal properties, Multi-scale modelling, Building energy.

## 1. Introduction

The rapid increase in world energy use has raised concerns about supply difficulties, depletion of energy resources, and serious environmental impacts such as ozone layer depletion, global warming, and climate change [1]. The International Energy Agency has collected dire data on trends in energy consumption about how greenhouse gas emissions will increase due to global warming, leading to extreme weather conditions around the world. Mitigating these challenges, the European Union (EU) aims to reduce carbon footprints by 80% to 95% below 1990 levels by 2050 [2]. Enhancing the energy performance of the building sector has been highlighted as a roadmap towards a competitive low-carbon economy by 2050, as 36% of total CO2 emissions come from this sector, where HVAC systems account for 50% total final energy consumption in the EU. Improving energy efficiency has become an increasingly popular research topic. The application of thermal energy storage, i.e., phase change materials (PCMs) in buildings has attracted many researchers in recent decades to improve building energy efficiency. Embedding PCMs in the building envelope/building materials will increase the latent heat storage capacity, thereby improving the indoor thermal environment and comfort [3]. Compared with other building materials used for thermal energy storage, PCM has so many advantages like high heat of fusion, high energy storage density, and constant phase change temperature.

However, PCMs also suffer from disadvantages, for instance, nonideal thermal conductivity and leakage during phase transitions. These disadvantages of PCM can be improved by preparing shape-stable polymer composites with PCM inclusions and embedding them in building components to improve the performance of building envelope. Microencapsulation of PCMs with polymeric shells stands out as one of the best confinement options for this application since it meets the requirements mentioned above and, additionally, polymers are cheap and have low density and thermal conductivity. Polyurethane (PU) rigid foams as the polymer matrix that can contain PCMs have been widely used for thermal insulation as the ultimate energy savers [4]. The air trapped within the honeycomb like structure develops passive insulation characteristics of foam in addition to polyurethanes' heat absorption capacity. The lowest thermal conductivity (between 0.02 and 0.05 W/mK), high mechanical and chemical stability at both high and low temperatures, the ability to form sandwich structures with various facer materials are their advantages. Compared with other insulating materials, PU-PCMs are highly competitive.

In recent years, many papers have studied the combination of PCM in PU matrix. Ana M. Borreguero et al. studied PU foams incorporating different percentages of microcapsules containing Rubitherm® RT27 [5]. Ming You et al. studied the thermal properties of The microencapsulated n-octadecane (MicroPCMs) with a styrene (St)–divinybenzene (DVB) co-polymer shell [6]. Ahmet Alper Aydın et al. showed the feasibility of PU–PCM composites by direct utilization of a PCM and its compatibility with the applied PU formulation recipe in terms of improved thermal characteristics [7]. Chunguang Yang et al. proved that the thermal energy PU storage–capacity PCM foam is enhanced significantly in their experimental study [8].

Most of these studies have focused on the synthesis methods and the thermal energy storage capacity of the composites, but few studies have addressed the thermal evaluation of PU-PCM foams in use. Based on the above situation, this paper focuses on the multi-scale modeling of PU-PCM at different scales, and conducts a case study to implement the impact of this composite on the energy efficiency of the building envelope.

This article is organized as follows. In the next section, we describe the multi-scale model for PU-PCM as well as the description of case study, which is based on the application of PU-PCM in building envelope. Subsequently, we discuss the results before we conclude our manuscript in last section.

## 2. Methods

In this section, we describe our hierarchical multi-scale modeling, which is inspired from the model presented in our previous research [9-14]. The models for PU-PCMs at different scales are based on finite element method - representative volume elements (FEM-RVEs) and can be reviewed in Fig. 1. For each scale, we first need to identify a suitable RVE size to subsequently upscale the material parameters to the next higher length scale. All relational effective parameters are summarized in Table 1. Later a case study is used to prove this application.

**Table 1**

*Relational effective parameters for each scale.*

| Effective scale | Effective length | Parameters |
|---|---|---|
| Meso | μm | Thermal conductivity of inclusion |
| | | Thermal conductivity of matrix |
| | | Interface resistance |
| | | Radius of inclusion |
| Macro | mm | Volume fraction |

**Figure 1**

*Multi-scale modelling scheme*

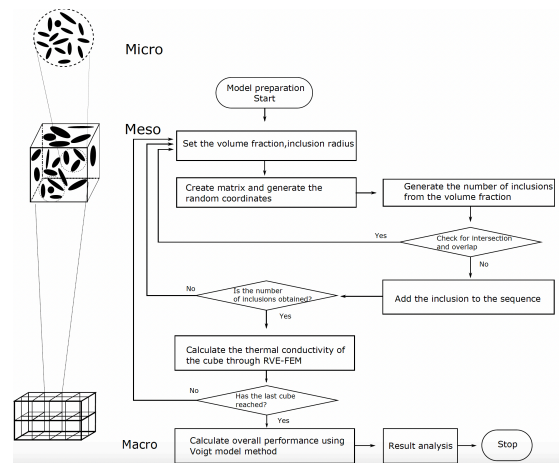

### 2.1 Multi-scale modelling of PU-PCMs

In our study there are mainly two different length scales - the meso-scale and the macroscopic scale. The key point is to link the material behaviour and engineering applications. A continuum model is employed at the mesoscale, which consists of the equivalent inclusions embedded in the polymer matrix. Based on the micromechanics approach, the volume elements should be sufficiently large to fully describe the statistics of the phase arrangement for the material we considered, i.e., they should be Representative Volume Elements (RVEs). In practice, smaller volume elements must typically be used due to limitations in available computational cost. So, the proper size of this RVE should be well defined and try to make it as small as possible. In this implementation a python script is applied to generate the microstructure of the RVE based on the algorithm which has been described in detail elsewhere. It avoids overlapping of the equivalent sphere and ensures the imposition of periodicity. The FEM package ABAQUS is run in all simulations. A typical discretization of the RVE with quadratic tetrahedra elements is illustrated in Fig.2.

**Figure 2**

*3-D cubic Representative Volume Element*

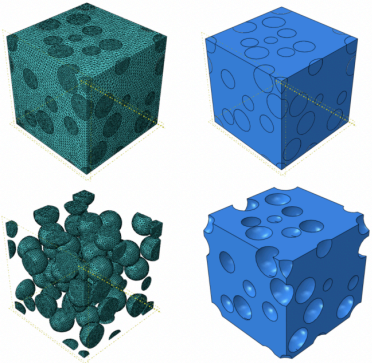

The field equation for heat conduction is given by:

$$C_f \frac{\partial \theta}{\partial t} + \nabla \cdot \boldsymbol{q} - Q = 0$$

where $\theta$ is temperature change, $\boldsymbol{q}$ the heat flux, $Q$ the body heat sourse, $C_f$ the heat capcity.

For quasi-steady problems, we have:

$$\nabla \cdot \boldsymbol{q} - Q = 0 \text{ in } \Omega$$

with natural boundary conditions:

$$\boldsymbol{q} \cdot \boldsymbol{n} = \bar{q} \text{ on } \Gamma_q$$

where $\boldsymbol{n}$ is the unit normal and $\bar{q}$ is the prescribed value of heat flux at $\Gamma_q$. The weak form for heat conduction is given by: Find $\theta \in \nu$ such that:

$$-\int \boldsymbol{q} \cdot \nabla(\delta\theta) d\Omega = -\int \bar{q} \delta\theta dS + \int Q \delta\theta d\Omega$$

Two different heat flux at both ends of the RVE, are applied, which is shown in Fig 3. It will generate a continuous heat flow through the entire RVE. The generated temperature gradient is necessary with Fourier's law to finally compute the macroscopic thermal conductivity:

$$\boldsymbol{q} = -\boldsymbol{\kappa} \cdot \nabla \theta$$

with

$$\boldsymbol{\kappa} = \begin{Bmatrix} k_{xx} & 0 & 0 \\ 0 & k_{yy} & 0 \\ 0 & 0 & k_{zz} \end{Bmatrix}$$

where $\boldsymbol{\kappa}$ is the thermal conductivity of the composite material. Regarding the boundary conditions at different RVE edges, the isotropic thermal conductivity with $\kappa_{xx} = \kappa_{yy} = \kappa_{zz}$ can be confirmed. The output of the meso-scale mode is the macroscopic thermal conductivity of the composite.

**Figure 3**

*Heat flux in Representative Volume Element*

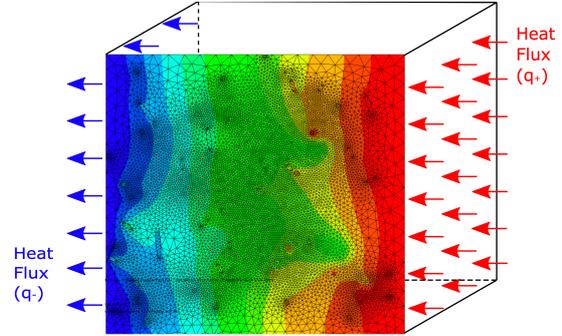

At the macroscopic scale, a larger homogenized structure is considered accounting for uncertainties. Therefore, cubic RVE with different thermal properties extracting from simulations at the meso-scale are randomly distributed in the macroscale, see Fig 3. Though it is principally possible to use FEM at the macro-scale, we employ the rule of mixture for computational efficiency, which is given by

$$\bar{X} = \frac{\sum_i X_i P_i}{\sum_i P_i}$$

where $X_i$ is the thermal properties of the $i$-th cubic simulation and $P_i$ is the weight.

## *2.2 PU-PCMs application: Case study*

We chose a single-family house that is common in European regions. The house is divided into two floors, and the layout of the house is as follows in Fig 4, Fig 5, and Fig 6 including a living room, the first bedroom on the first floor, the second bedroom on the first floor, stairwell, bathroom, storage room and the third bedroom (second floor), the fourth bedroom (second floor). The specific location of the house is Umeå Sweden (63°49′30″N 20°15′50″E), where it is a typical city in northeast Sweden, the capital of Västerbotten County. Umeå has a subarctic climate (Dfc), bordering on a humid continental climate (Dfb) with short and fairly warm summers. Winters are lengthy and freezing but usually milder than in areas at the same latitude with a more continental climate. Average January temperature is about −7 °C (19 °F), July is 16 °C (61 °F).

Based on the above location information and climate distribution, in this case study, the main consideration is the energy consumption for heating houses in the Umea area in summer. The application of PU-PCM can store and release energy through phase change while reducing the U-value of the house. In this scenario, we choose to add an additional layer of PU-PCM on the interior walls and ceiling of the house. The engineering parameters of PU-PCMs are conveyed by previous multiscale modeling. Our simulation is mainly based on the hourly trend of temperature in Umeå in 2022 as the external temperature load. The change of indoor temperature is mainly the response of the external temperature change and fuel heating.

**Figure 4**

*The application of PU-PCMs in building envelope*

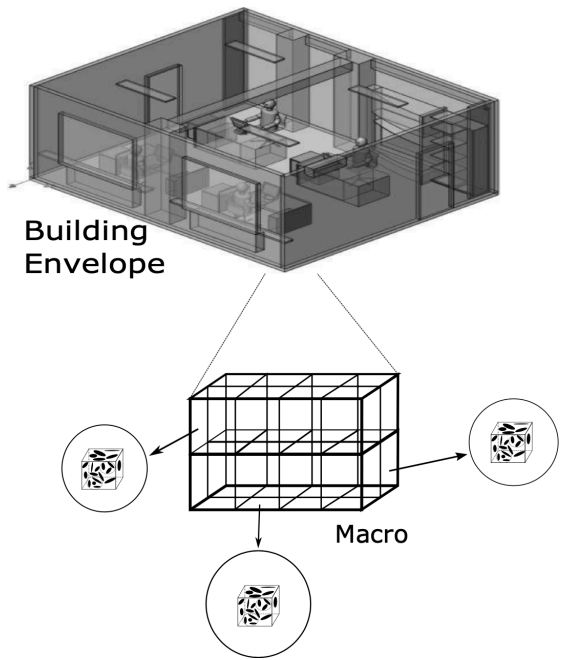

**Figure 5**

*Building Exterior for Case Study*

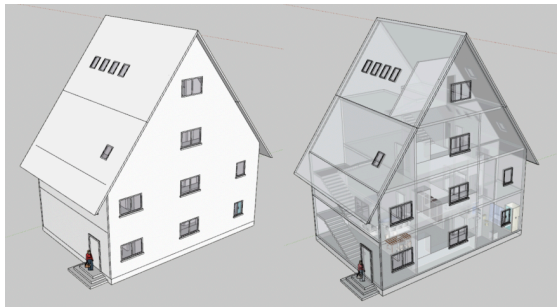

## 3. Results and Discussion

In this section we mainly introduce the engineering parameters of PU-PCMs composites and their performance in this case study application.

As shown in the Fig. 7 below, this is the result of RVE based on FEM simulation. After applying a thermal load in the corresponding direction, we can obtain the temperature distribution of its response. Through Fourier's law, the effective thermal conductivity of PU-PCMs composites can be computed, and the result is shown in Table 2.

**Figure 6**

*Architectural plans for singe-family house*

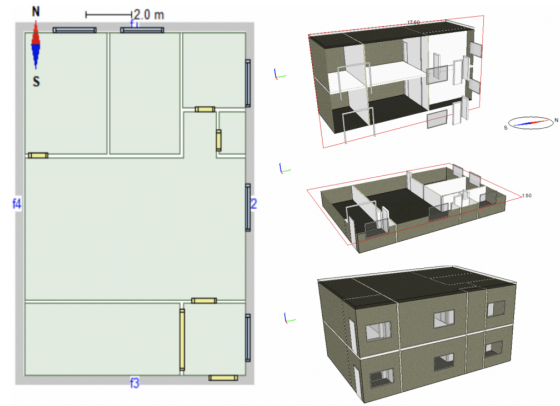

**Figure 7**

*The temperature distribution of PU-PCMs RVE*

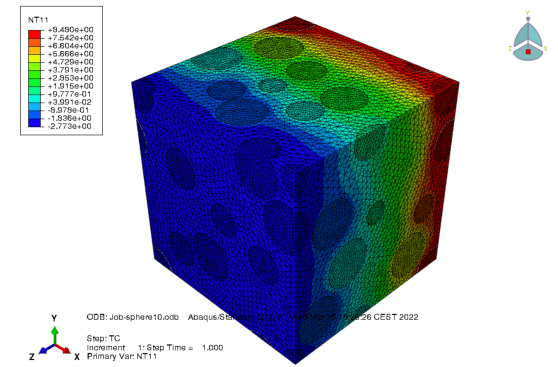

Then we import the obtained engineering parameters of PU-PCMs as modeling variables into the application of single-family houses. Through partial differential equations, a physical simulation of the entire house is established. We use Runge-Kutta methods to obtain the annual energy consumption for the whole year of 2022 and the hourly indoor temperature change throughout the year. The annual energy consumption results are shown in the table 3 and table 4. We calculate two comparative models without and with PU-PCMs interlayer as part of the building envelope. From the results, it can be seen that PU-PCMs can provide significant energy savings for single-family houses in Umeå in 2022 weather conditions. Compared with the predicted data, it can be found that the energy efficiency has been improved by 2.6% in this case study, which means the which means that there will be a corresponding reduction in $CO_2$ emissions at the same time.

According to Fig. 8 and Fig. 9, the daily energy usage distribution, we can also find that PU-PCMs do not directly change the peak range of energy consumption, but only improve energy efficiency.

**Table 2**

*Thermal conductivity of PU-PCMs*

| Effective scale | Effective length | Parameters | Value |
|---|---|---|---|
| Meso | μm | Thermal conductivity of inclusion | 0.56 W/m·K |
| | | Thermal conductivity of matrix | 0.036 W/m·K |
| | | Interface resistance | $35\ MW\ m^2/K$ |
| Macro | mm | Volume fraction | 20% |
| Macro | mm | Effective thermal conductivity | 0.24 W/m·K |

**Table 3**

*Annual Energy usage without PU-PCMs*

| | | Purchased energy (kWh) | (kWh/m2) |
|---|---|---|---|
| 🟨 | Lighting, facility | 32199 | 126.1 |
| 🟦 | Electric cooling | 16062 | 62.9 |
| 🟦 | HVAC aux | 7258 | 28.4 |
| 🟥 | Fuel heating | 15369 | 60.2 |
| | Total, Facility electric | 55519 | 217.4 |
| | Total, Facility fuel | 15369 | 60.2 |
| | Total | 70888 | 277.5 |
| | Equipment, tenant | 24149 | 94.5 |
| | Total, Tenant electric | 24149 | 94.5 |
| | Grand total | 95037 | 372.1 |

**Table 4**

*Annual Energy usage with PU-PCMs in Building envelope*

| | | Purchased energy (kWh) | (kWh/m2) |
|---|---|---|---|
| 🟨 | Lighting, facility | 32199 | 126.1 |
| 🟦 | Electric cooling | 16322 | 64 |
| 🟦 | HVAC aux | 7258 | 28.4 |
| 🟥 | Fuel heating | 14973 | 58.6 |
| | Total, Facility electric | 55809 | 218.5 |
| | Total, Facility fuel | 14973 | 58.6 |
| | Total | 70782 | 277.1 |
| | Equipment, tenant | 24149 | 94.5 |
| | Total, Tenant electric | 24149 | 94.5 |
| | Grand total | 94931 | 371.7 |

After obtaining the annual energy consumption data, we also consider thermal comfort. We have chosen the first bedroom in this single-family house as an example. We mainly consider the real-time temperature for each of the 8760 hours of the year. We also define an optimal comfort temperature of 21-25 degrees Celsius, and a fluctuation of 1 degree Celsius within this range is considered a good comfort temperature. 18-20 degrees Celsius is defined as an acceptable temperature. Below 18 degrees Celsius and above 26 degrees Celsius are defined as unacceptable temperature range. So our thermal comfort simulation is also based on the above definition interval. As shown in Figs. 10 and 11, we can observe that according to the above definition, the addition of PU-PCMs layer in the building envelope can significantly improve the thermal comfort time, respectively increasing the best thermal comfort time by 39.17%, 17.56% improvement of good thermal comfort time and 8.48% improvement of acceptable thermal comfort range, meanwhile, unacceptable time decreased by 3.79%. These data and those figures show that adding PU-PCMs can significantly improve the thermal comfort in various time periods of a year, especially in summer.

**Figure 8**

*Daily Energy usage plot without PU-PCMs*

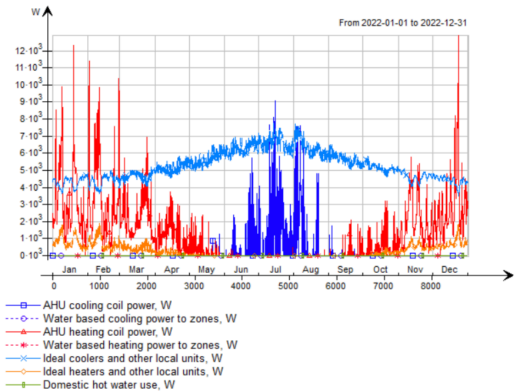

**Figure 9**

*Daily Energy usage with PU-PCMs in Building envelope*

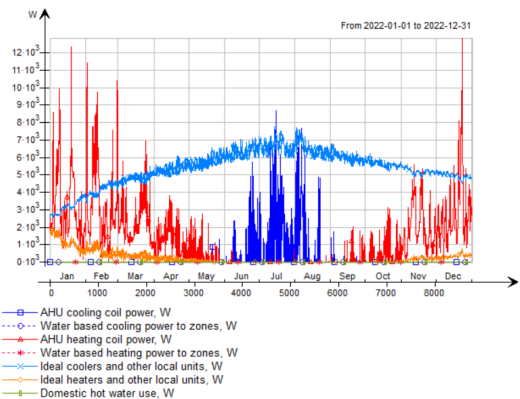

**Figure 10**

*Annual thermal comfort hours in 1st bedroom*

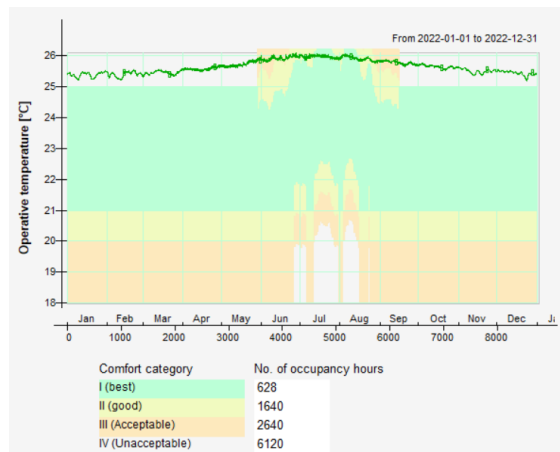

**Figure 11**

*Annual thermal comfort hours in 1st bedroom with PU-PCMs enhanced*

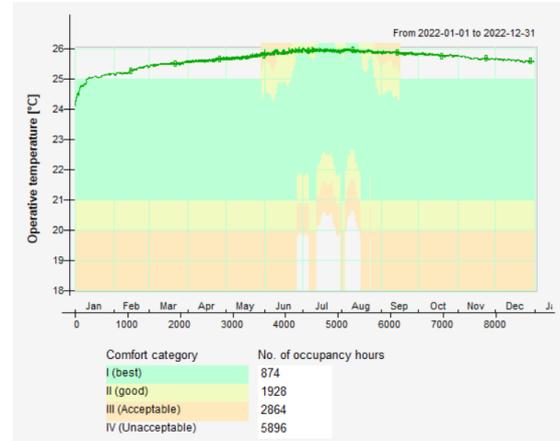

According to the above data, tables, and figures, it can be seen directly that the application of PU-PCMs as an additional layer in the building envelope can not only improve energy efficiency, but also significantly improve the thermal comfort of the indoor environment. It is also evident in this case study based on the Umeå region in 2022 that, regardless of cooling, the application of this composite material can effectively improve the indoor thermal comfort period, especially during the hot summer period.

## 4. Conclusion

In this study, we develop a hierarchical multi-scale model of a PU-PCM foam composite and study the thermal conductivity of this material at the meso-scale and macro-scale. Subsequently, the effective engineering parameters of these materials can be computed by the RVE-finite element method. We later conducted a case study based on the performance of this designed material to fully consider the thermal comfort of the occupants of a building envelope with the application of PU-PCMs composite in a single house. At the same time, we also simulate the annual energy consumption of this case. All the outcomes show that this design is promising, enabling a Passive design of building energy and significantly improving occupants' comfort. In summary, the following conclusion can be summarized:

1. The energy efficiency has been improved by 2.6% in the case study in the Umeå region in 2022. Meanwhile a corresponding reduction in $CO_2$ emissions can be deducted.

2. the addition of PU-PCMs layer in the building envelope can significantly improve the thermal comfort time, respectively increasing the best thermal comfort time by 39.17%, 17.56% improvement of good thermal comfort time and

   8.48% improvement of acceptable thermal comfort range, meanwhile, unacceptable time decreased by 3.79%.

3. In areas where substantial summer cooling is not considered, such as Umeå, the application of this composite material can effectively improve the indoor thermal comfort period, especially during the hot summer period.

## 5. Acknowledgements

We gratefully acknowledge the support of the Kempe Foundation Sweden (Kempestiftelserna - Stiftelserna J.C. Kempes och Seth M. Kempes minne) and EU project H2020-AURORAL (Architecture for Unified Regional and Open digital ecosystems for Smart Communities and Rural Areas Large scale application).

We also gratefully acknowledge the support of The Arctic Centre's strategic funds for research trips and conference attendances.

The computations handling were enabled by resources provided by the Swedish National Infrastructure for Computing (SNIC) at High Performance Computing Center North (HPC2N) partially funded by the Swedish Research Council through grant agreement no. 2018-05973.